\documentclass[a4paper,twocolumn, nofootinbib,superscriptaddress,aps,prl, 14pt,eqsecnum,notitlepage,showkeys]{revtex4-1}
\usepackage{multirow}
\usepackage{graphicx}
\usepackage{bm}
\usepackage{dcolumn}
\usepackage{hyperref}
\usepackage{gensymb}
\usepackage{amsmath}
\usepackage{dcolumn}    % Align table columns on decimal point
\usepackage{ifpdf}
\usepackage{amssymb,lineno,amsfonts}
\usepackage{graphicx}   % Include figure files
\usepackage{bm}         % bold math
\usepackage{bbm}
\usepackage{mathrsfs}
\usepackage{upgreek}
\usepackage{mathtools}
\usepackage{epstopdf}
\usepackage{setspace}
\usepackage{hyperref}
\usepackage{natbib}
\usepackage{esvect}
\usepackage{amsmath}
\usepackage[usenames,dvipsnames]{xcolor}
\definecolor{med-blue}{RGB}{25,25,112}
\hypersetup{colorlinks, linkcolor={blue},citecolor={blue}, urlcolor={blue}}
\usepackage{times }
\usepackage [english]{babel}
\usepackage [autostyle, english = american]{csquotes}
\MakeOuterQuote{"}
\begin{document}
\title{Improper multiferroicity and colossal dielectric constants in Bi$_{2}$CuO$_{4}$}
\author{Jitender Kumar}
\affiliation{Department of Physics, Indian Institute of Science Education and Research, Dr. Homi Bhabha Road, Pune 411008, India}
\author{Soumendra Nath Panja}
\affiliation{Department of Physics, Indian Institute of Science Education and Research, Dr. Homi Bhabha Road, Pune 411008, India}
\author{Amit Kumar Naiya}
\affiliation{UGC-DAE Consortium for Scientific Research, University Campus, Khandwa Road, Indore, India}
\author{Sunil Nair}
\affiliation{Department of Physics, Indian Institute of Science Education and Research, Dr. Homi Bhabha Road, Pune 411008, India}
\affiliation{Centre for Energy Science, Indian Institute of Science Education and Research, Dr. Homi Bhabha Road, Pune 411008, India}
\date{\today}
\begin{abstract} 
The layered cuprate Bi$_{2}$CuO$_{4}$ is investigated using magnetic, dielectric and pyroelectric measurements. This system is observed to be an improper multiferroic, with a robust ferroelectric state being established near the magnetic transition. Magnetic and dielectric measurements indicate the presence of a region above the antiferromagnetic Neel temperature with concomitant polar and magnetic short range order. Bi$_{2}$CuO$_{4}$ is also seen to exhibit colossal dielectric constants at higher temperatures with clearly distinguishable grain and grain boundary contributions, both of which exhibit non-Debye relaxation. 
\end{abstract}
\pacs{Pacs}
\maketitle
\section{Introduction}

The discovery of high temperature superconductivity in electron and hole doped cuprates \cite{Bednorz1986,Tokura} have ensured that they have remained at the forefront of experimental and theoretical investigations in condensed matter physics. However, unconventional superconductivity is not the only exotic phenomena discovered in this class of materials. For instance, the under-doped cuprate La$_{2}$CuO$_{4+x}$ was reported to exhibit a low temperature ferroelectric state arising as a consequence of a distortion in the CuO$_{6}$  octahedra, with an associated magnetoelectric coupling being tuned through the antisymmetric Dzyaloshinskii-Moriya interaction \cite{cite3,cite4}. As is well known, the magnetoelectric multiferroics refer to a class of materials where magnetic and polar orders co-exist, and such materials are in vogue for the rich physics associated with these novel states as well as potential device applications \cite{Eerenstein,Mostovoy,KFWang}. Of special interest are systems known as improper (or Type-II) multiferroics, where ferroelectricity arises as a direct consequence of magnetic order \cite{KimuraNature,Lawes,kimura,kimuraPrl,Yamasaki,Taniguchi}. 

Layered cuprates of the form $R_{2}$CuO$_{4}$ are known to crystallize in three different structural motifs, typically dictated by the ionic radius of the $R$ site ion. When $R=$ La, a highly distorted network of corner sharing CuO${_6}$ octahedra is observed (called the $T$-phase), whereas systems with smaller $R$ site ions (Nd, Sm or Pr) prefer a corner shared network of square planar CuO${_4}$ units (called the $T'$-phase)\cite{Tokura}. Some systems which constitute of both large and small $R$ site ions are seen to crystallize in a phase which is made up of half a unit cell each of the $T$ and $T'$ phases, and are typically referred to as the $T^*$ phase in cuprate literature \cite{Tokura,sawa,cite5,cite6}. Interestingly, the closely related Bi${_2}$CuO${_4}$ system exhibits a very different structure, though the ionic radii of Bi${^{3+}}$ (1.31\AA) is very similar to that of La${^{3+}}$ (1.30\AA). This is presumably due to the strongly covalent nature of the Bi-O bond, which reduces the effective co-ordination of the $R$ site from 9 to 6 \cite{cite8,cite10}. Bi${_2}$CuO${_4}$ is reported to crystallize in a tetragonal symmetry (space group $P4/ncc$) and is now established to be a 3D antiferromagnet, with a transition temperature ($T{_N}$) $\approx$ 50 K \cite{cite8,cite9}. As is shown in Fig.1, the structure of Bi${_2}$CuO${_4}$  is characterized by the presence of seemingly isolated CuO${_4}{^{6-}}$ square plaquettes which are staggered in a chain-like fashion along the crystallographic $z$ axis. Unlike other members of the $R{_2}$CuO${_4}$ family, where the magnetism is driven by conventional Cu-O-Cu superexchange and Cu-O-O-Cu extended superexchange pathways\cite{bush}, it has been suggested that in Bi${_2}$CuO${_4}$, the magnetism is driven by four dissimilar Cu-O-Bi-O-Cu paths \cite{cite8,cite13,cite14}. The magnetic structure has also been determined using neutron diffraction, and is thought to be made up of ferromagnetically ordered spins along the $c$ axis, with these chains being  antiferromagnetically coupled to the neighbouring ones\cite{cite8,paul}. Though a preliminary report exists on the dielectric properties of this system \cite{cite15}, the behavior of the dielectric properties in the vicinity of the magnetic transition remains to be investigated. The discovery of multiferroicity in the layered cuprates clearly warrants a more serious investigation of Bi${_2}$CuO${_4}$, especially since the presence of the 6$s{^2}$ lone pair on the Bi${^{3+}}$ ion could offer an additional means of inducing ferroelectric polarization in this system akin to that known in the perovskite BiFeO${_3}$ \cite{cite12}. 
\begin{figure}
 	\centering
 	\hspace{-0.5cm}
 	\includegraphics[scale=0.5]{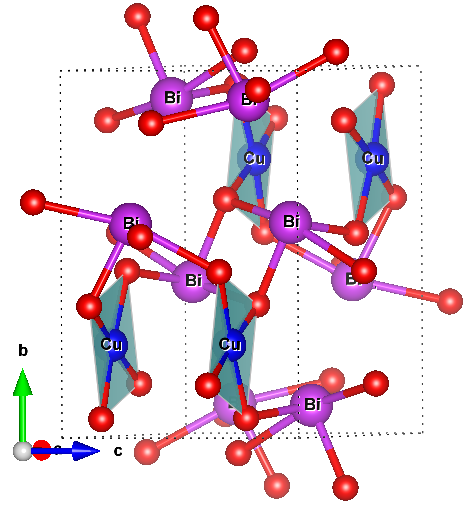}
 	\caption{The crystal structure of Bi$_2$CuO$_4$ comprising of CuO${_4}{^{6-}}$ square plaquettes staggered in chains along the crystallographic $c$ axis.}
 	\label{Fig1}
 \end{figure}
Here, we investigate the magnetic, dielectric and pyroelectric properties of Bi${_2}$CuO${_4}$, and observe that this system is a hitherto undiscovered multiferroic, with a ferroelectric polarization being set up in close vicinity to the magnetic phase transition. At temperatures well above the magnetic ordering temperature ($T{_N}$), this system exhibits a colossal dielectric behavior, and two distinct relaxation peaks in the dielectric loss arising from the grain and the grain boundary response can be clearly identified. 
\begin{figure}
	\centering
		\hspace{-0.6cm}
	\includegraphics[scale=0.40]{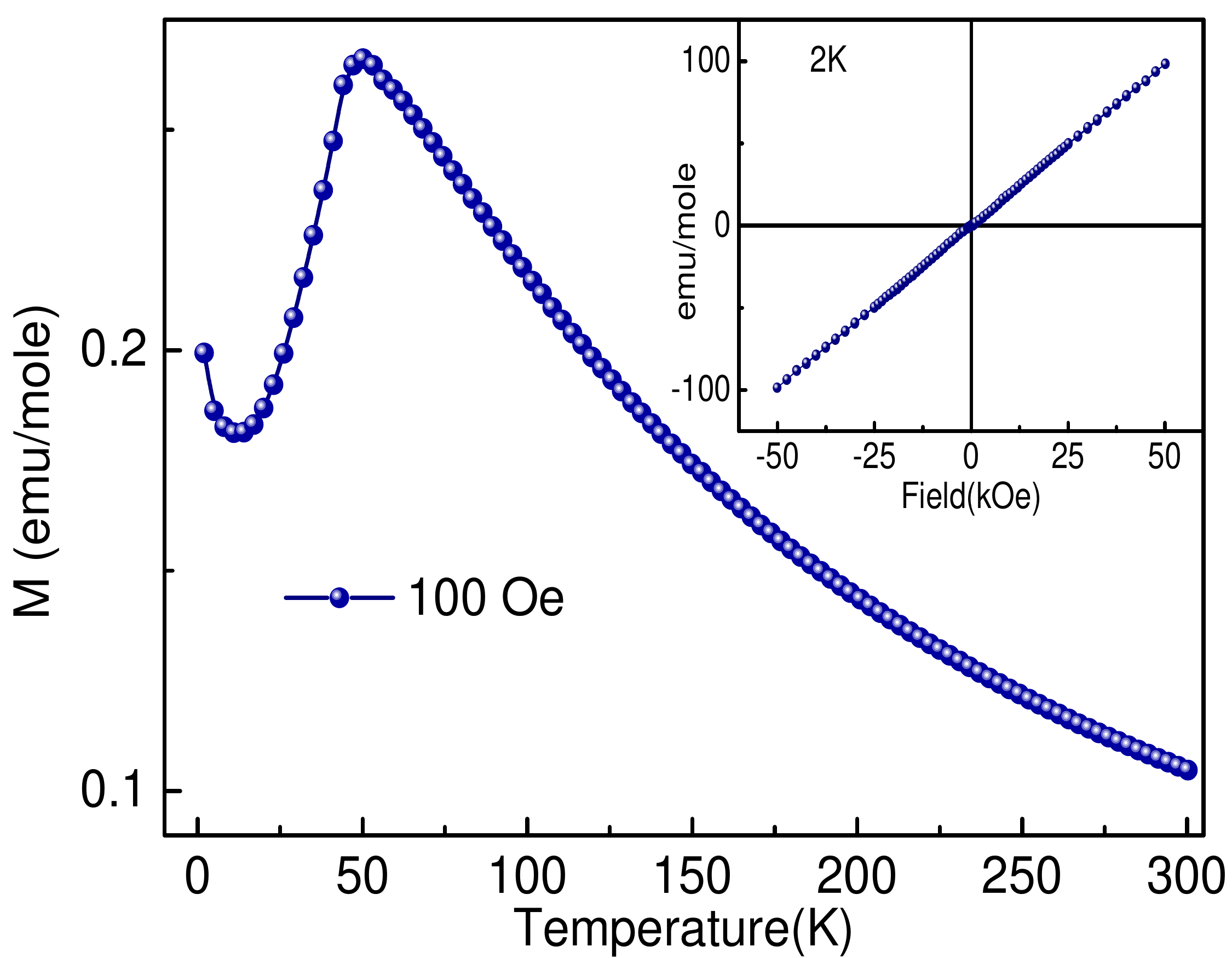}
	\caption{The dc Magnetization of Bi$_2$CuO$_4$ measured as a function of temperature in the Zero Field Cooled (ZFC) protocol. The inset depicts an $MH$ isotherm as measured at 2 K.}
	\label{Fig2}
\end{figure}
\section{Experimental}
Polycrystalline specimens of Bi$_2$CuO$_4$ were prepared by the conventional solid state reaction method by using Bi$_2$O$_3$ and CuO as the ingredients. Stoichiometric amounts of these oxides were mixed and heated at 700$^o$C in air for 24 hours, following which they were pelletised and reheated at 780$^o$ for another 24 hours. Powder X-ray diffraction data was recorded using a Bruker D8 Advance diffractometer with CuK$_{\alpha}$ source under the continuous scanning mode. Magnetization measurements were performed using a Quantum Design (MPMS-XL) SQUID magnetometer. Temperature dependent dielectric measurements were performed by using Alpha-A High Performance Frequency Analyzer from Novocontrol Technologies. The dielectric measurements under magnetic field were performed by using the Manual Insertion Utility Probe of the MPMS-XL magnetometer. During the dielectric measurements, 1 Volt excitation signal was used to probe the dielectric response. The pyroelectric measurements were performed using a Keithley Sourcemeter (Model 2612B) and a Picoammeter (Model 6482) and the ferroelectric polarization was derived from the pyroelectric current by integrating over time. 
\section{Results and Discussions}
\begin{figure}
	\centering
	\hspace{-0.5cm}
	\includegraphics[scale=0.40]{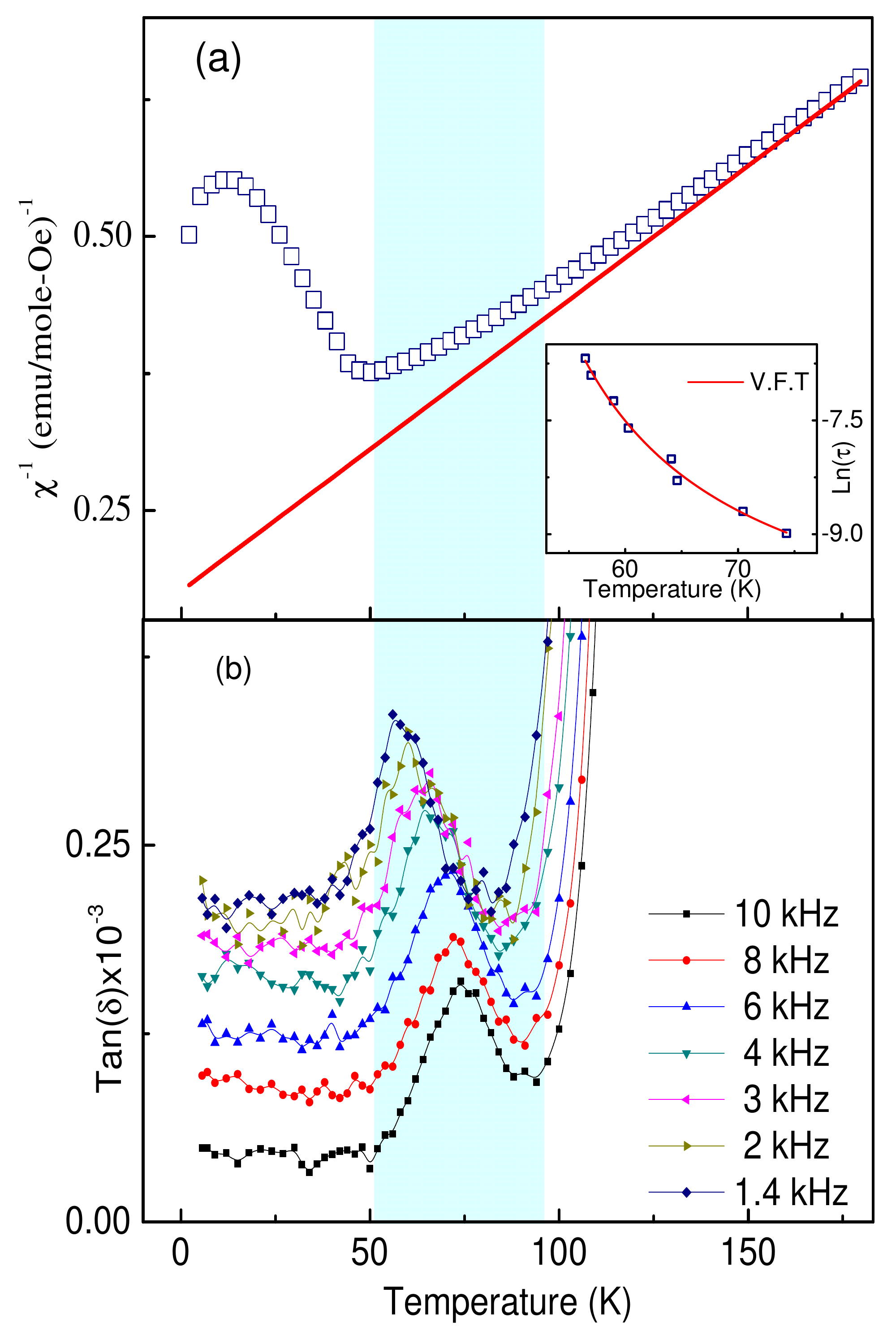}
	\caption{(a) depicts a view of the inverse dc magnetic susceptibility exhibiting deviation from a Curie Weiss fit in the region above the magnetic ordering temperature.  (b) shows the dielectric loss tangent as a function of temperature at various probed frequencies depicting a frequency and temperature dependent relaxation above $T_N$. The inset of (a) shows the activated VFT behavior of this relaxation process.}
	\label{Fig3}
\end{figure}	
\begin{figure}
	\centering
	\hspace{-0.5cm}
	\includegraphics[scale=0.27]{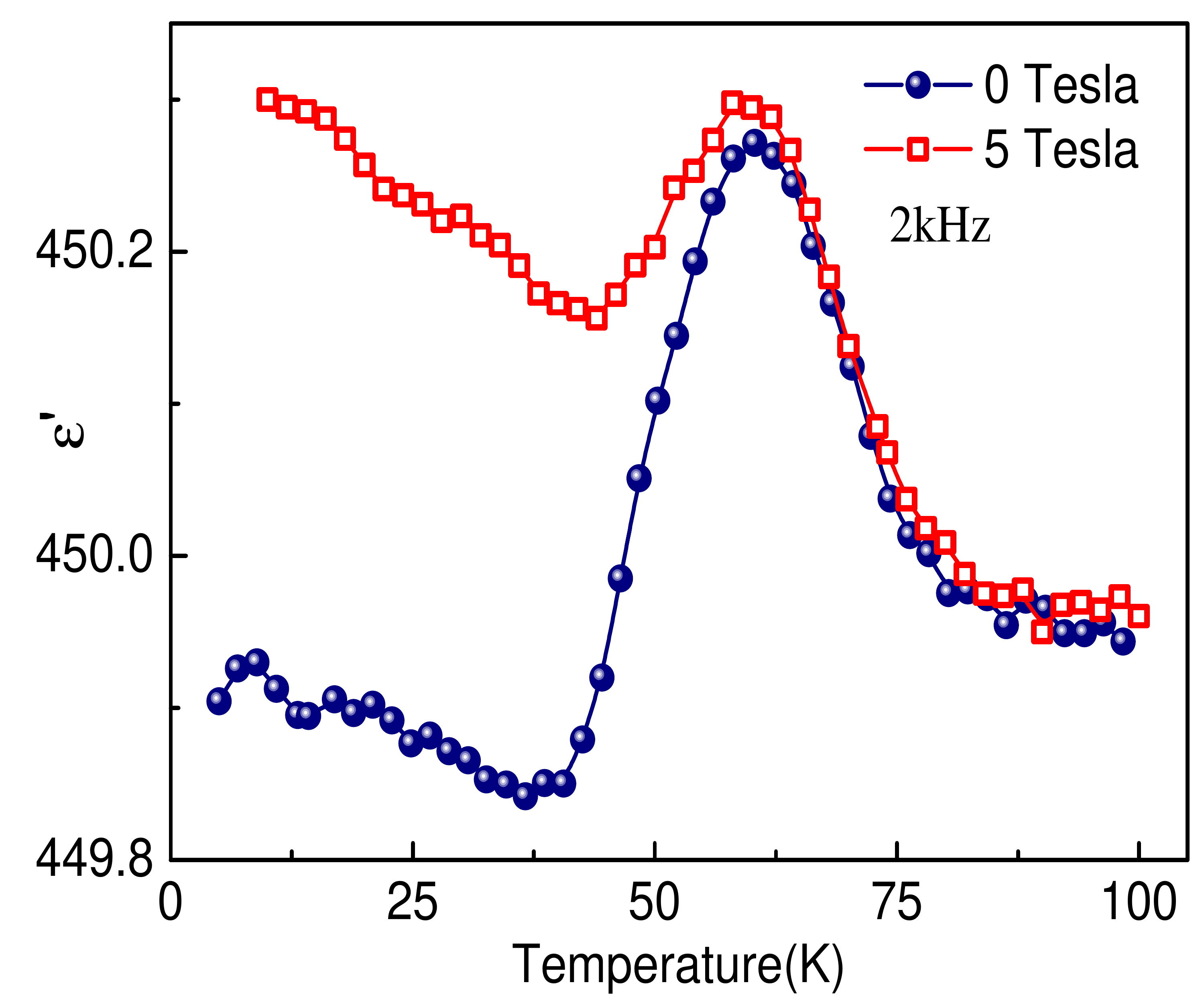}
	\caption{The temperature dependence of the real part of the dielectric permittivity ($\epsilon'$) as measured at zero and 5 Tesla, indicating that within the magnetically ordered state, $\epsilon'$($T$) can be tuned by an applied magnetic field.}
	\label{Fig4}
\end{figure}
The dc magnetisation of Bi$_2$CuO$_4$ as measured in the Zero Field Cooled (ZFC) protocol is depicted in Fig. \ref{Fig2}. A clear antiferromagnetic-like transition is observed at $\approx$ 50 K, in good agreement with previous reports \cite{cite15}. The magnetic transition temperature of Bi$_2$CuO$_4$ is smaller than that of its La and Nd based counterparts, presumably as a consequence of the larger exchange path which drives magnetic long range order. The inset of Fig. \ref{Fig2} shows a $MH$ isotherm as measured at 2 K, which is typical of a robust antiferromagnet, with no sign of any saturation up to the highest applied field of 5 Tesla. At temperatures in excess of 150 K, the inverse of the susceptibility follows a Curie-Weiss fit, giving $\theta$ = -68.4 K and an effective magnetic moment of $\mu_{eff}$ = 1.76 $\pm$ 0.01 $\mu_{B}$, which matches very well with the spin-only value (1.73$\mu_{B}$) expected from a Cu$^{2+}$system. Deviation from linearity extends to 2$\times$$T_N$, indicating that short range correlations persist well above the transition temperature as is shown in Fig. \ref{Fig3}a. 

Interestingly, the influence of these correlations is also evident in the dielectric measurements, and the loss tangent exhibits a clear relaxation peak in this regime (Fig. \ref{Fig3}b). The temperature at which the peak is observed shifts to lower temperatures with decreasing measurement frequency, which is a typical signature of dielectric relaxation in polar glasses or relaxor ferroelectrics \cite{cite18}, with the relaxation time following a Vogel-Fulcher-Tamman (VFT) behavior given by $\tau  = {\tau_0}  \exp  \frac{E}{K_B(T-T_f)}$. Here $\tau$ is the relaxation time, $\tau_0$ is the pre-exponential factor, $E$ is the activation energy of the process and $K_B$ is the Boltzmann constant. $T$ refers to the temperature at which the maxima is observed, and $T_f$ is the static freezing temperature where the dynamics of the relaxing entities become frozen. As is shown in the inset of Fig. \ref{Fig3}a, this relaxation process conforms to the VFT model, with an activation energy of 5.52 $\pm$ 2.06 meV, and a $T_f$ of 41.2 $\pm$ 3.7 K. This feature can originate from polar nano regions (PNR) coupled to the short ranged magnetic correlations, and could be looked upon as a possible precursor to the onset of long range ferroelectric order. Interestingly, we do not observe any signature of this relaxation in the real permittivity, presumably due to a small change in the dielectric constant as a function of frequency. However, a prominent peak is observed in the real part of the the dielectric susceptibility $\epsilon'$($T$) in the vicinity of the magnetic transition, indicating a coupling of the electric and magnetic order parameters in this system (Fig. \ref{Fig4}). Moreover, within the magnetically ordered state, $\epsilon'$($T$) also exhibits a clear magnetic field dependence, indicating that the polar state can be tuned as a function of the applied magnetic field. We note that the change observed in the dielectric constant on applying a magnetic field is not very large, and is of the order of $\approx$ 0.1 \% at $T=$ 10 K and $H=$ 5 Tesla.   
\begin{figure}
	\centering
	\hspace{-0.5cm}
	\includegraphics[scale=0.36]{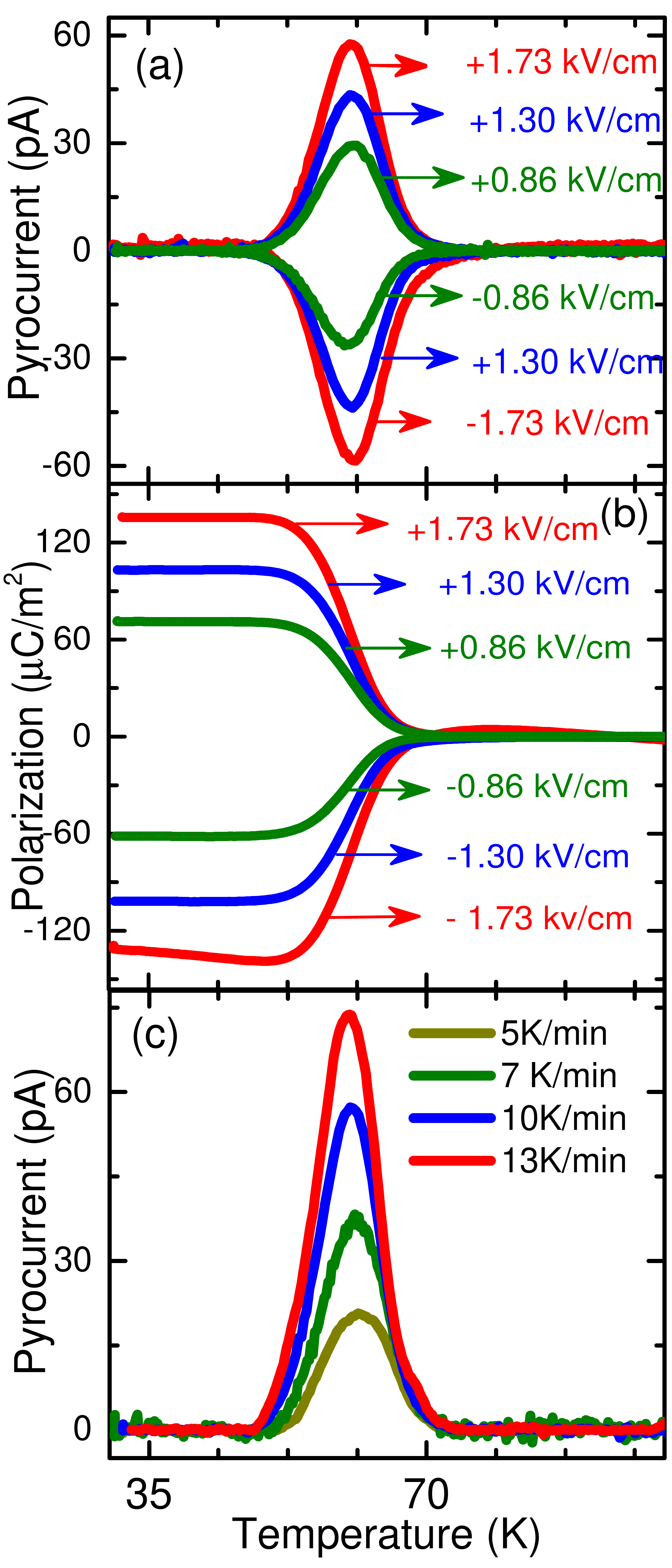}
	\caption{(a) shows the pyrocurrent as measured with a warming rate of 10 K/min at different poling fields. The reversal of the pyrocurrent on reversal of the poling field direction denotes a true ferroelectric state. (b) depicts the polarization as obtained from integrating the measured pyrocurrent. (c) shows the pyrocurrent as measured at different warming rates with the poling field being kept constant at 1.73 kV/cm.}
	\label{Fig5}
\end{figure}
\begin{figure}
	\centering
	\hspace{-0.5cm}
	\includegraphics[scale=0.32]{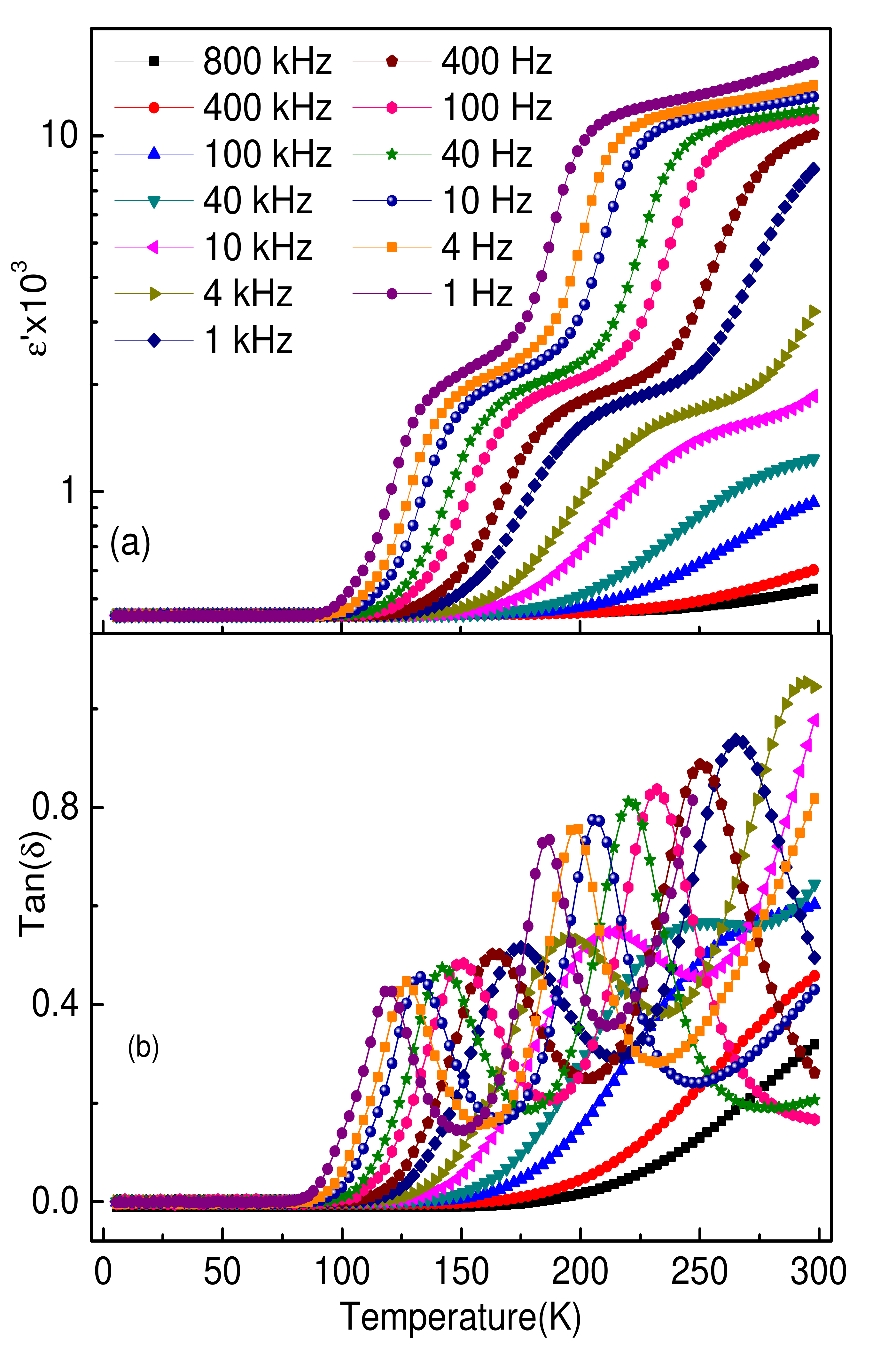}
	\caption{The real part of the dielectric permittivity (a) and the loss tangent (b) plotted as a function of temperature at different probed frequencies. }
	\label{Fig6}
\end{figure}

Confirmation of the multiferroic nature of Bi$_2$CuO$_4$ is provided by temperature dependent measurements of the pyroelectric current, as is shown in Fig. \ref{Fig5}. These measurements were done in the standard parallel plate geometry with different poling fields, and the pyroelectric current was measured at a warming rate of 10 K/min (Fig. \ref{Fig5}a). A peak in the pyroelectric current was detected in close proximity to the magnetic ordering temperature, with its sign being flipped on reversing the direction of the applied electric field, thus signifying a true ferroelectric state. The effective ferroelectric polarization values were deduced to be of the order of 120 $\mu C/m{^2}$ at a poling field of 1.73 kV/cm as is shown in Fig. \ref{Fig5}b. These measurements were also validated by repeating the measurement with different warming rates (Fig. \ref{Fig5}c), with the invariance of the ferroelectric transition temperature ruling out possible experimental artifacts \cite{kohara}. 
\begin{figure}
 	\centering
 	\hspace{-0.5cm}
 	\includegraphics[scale=0.45]{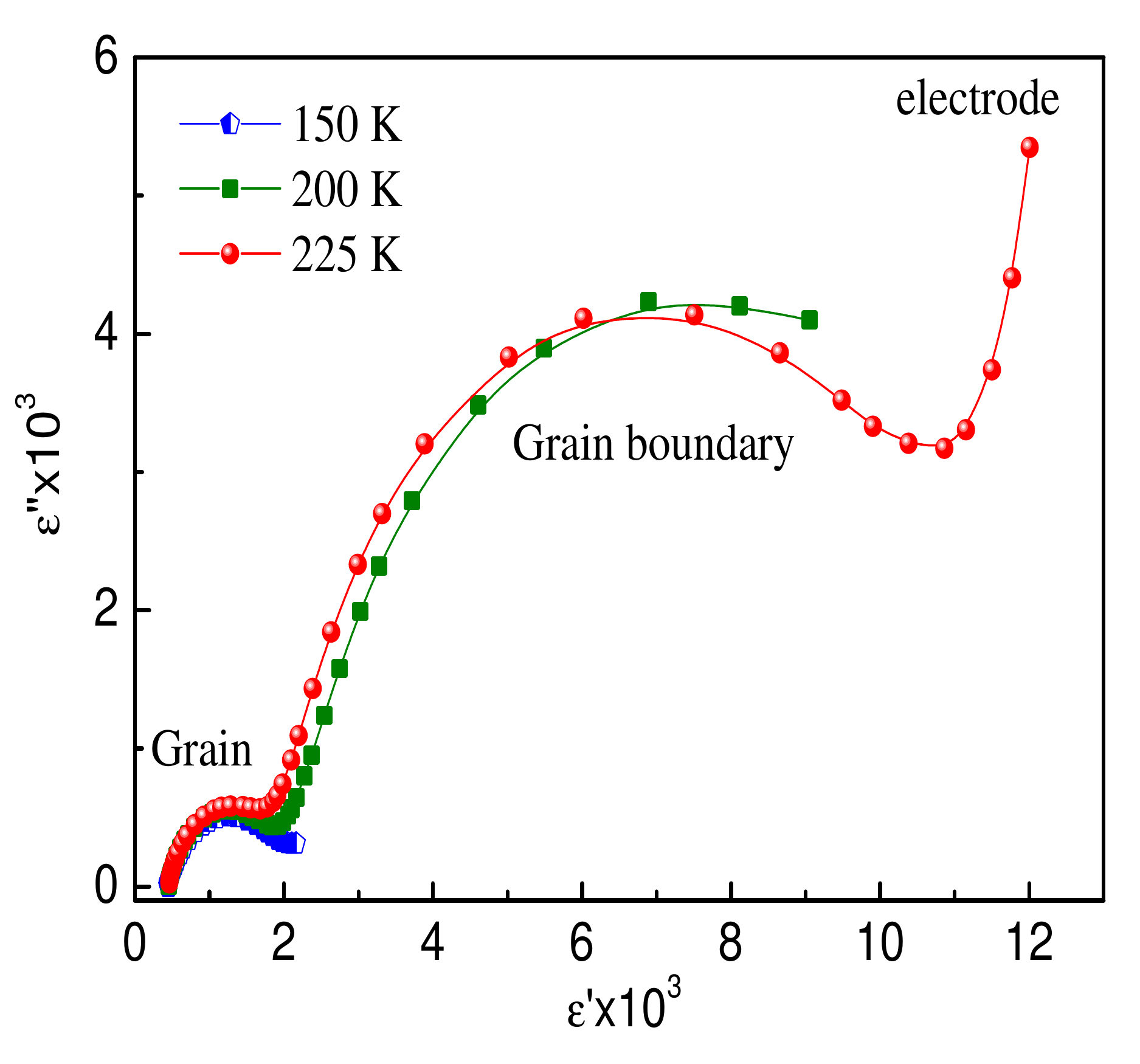}
 	\caption{Cole-Cole plots of Bi$_2$CuO$_4$ as measured at 150 K, 200 K and 225 K, with the grain, grain-boundary, and electrode contributions being clearly discernible. }
 	\label{Fig7}
 \end{figure}

Magnetically induced ferroelectric order can either arise from the antisymmetric exchange interaction as in the case of spiral magnets, or from the symmetric exchange striction as in the case of collinear magnets. The magnetic order in Bi$_2$CuO$_4$ is reported to be collinear, with spins being aligned along the crystallographic $c$ axis with ferromagnetic intra-chain and antiferromagnetic inter-chain interactions. Our dc susceptibility data also shows that the obtained effective magnetic moment per Cu site (1.76 $\pm$ 0.01$\mu_{B}$) is basically the spin-only moment (1.73$\mu_{B}$), thus indicating that the orbital contribution to the magnetic moment arising from the spin orbit coupling is negligible. Moreover, an early experimental report \cite{paul} clearly shows that Bi$_2$CuO$_4$ exhibits a hysteretic exchange-striction driven discontinuity in both the $a$ and $c$ lattice parameters across the magnetic phase transition, though the symmetry of the crystallographic unit cell remains unchanged. All these factors point towards  Bi$_2$CuO$_4$ being an exchange striction driven improper multiferroic. The values of ferroelectric polarization in this system are also comparable to that observed in polycrystalline specimens of other exchange striction multiferroics like CdV$_2$O$_7$ and HoMnO$_3$ \cite{cvo,lorenz}. We note that polarization measurements performed in polycrystalline specimens at moderate poling fields do tend to underestimate the true polarization value associated with the ferroelectric state. Though the strongly covalent Bi-O bond would be expected to be influenced by the exchange striction driven structural distortion, the data at our disposal does not indicate an appreciable contribution from the steriochemical activity of the Bi${^{3+}}$ lone pair to the effective polarization. For instance, polarization values measured in polycrystalline specimens of the lone-pair multiferroic BiFeO$_3$ \cite{bfo} are more than two orders of magnitude higher than that measured in Bi$_2$CuO$_4$. However, measurements on single crystalline specimens and additional theoretical inputs would be needed to conclusively elaborate on the role of the Bi${^{3+}}$ lone pair in this ferroelectric state. 
 
With the multiferroic nature of Bi$_2$CuO$_4$ having been established, we now focus on the high temperature dielectric properties of this system. There have been reports of large dielectric constant values in related cuprate systems \cite{die}, though an earlier report on Bi$_2$CuO$_4$ reported rather modest values ($100-500$) of the dielectric constant at room temperatures. Fig. \ref{Fig6} shows the dielectric constant and loss tangent measured as a function of temperature at different frequencies. The room temperature value of dielectric constant in our Bi$_2$CuO$_4$ specimen is of the same order as that reported for other cuprates \cite{cite22,cite23} and substantially higher than that reported for this system earlier \cite{cite15}. $\epsilon^{'}$($T$) exhibits two steps from low-temperature values ($\approx400$) to intermediate higher values ($\approx$2000) succeeded by another step up-to colossal values of the order of 10$^4$. Correspondingly, the loss tangent shows two distinct relaxation peaks, both of which shift towards higher temperatures on increasing the measurement frequencies (Fig. \ref{Fig6}b). These are features typical of materials exhibiting Colossal Dielectric Constants (CDC), with the cubic CaCu$_{3}$Ti$_{4}$O$_{12}$ being a popular example \cite{cite24, cite25}. Current wisdom is that this behavior can arise as a consequence of the interfaces between grain and grain boundaries which act as parallel plate capacitors with large dielectric constants. Originally proposed by Maxwell and Wagner in the context of heterogeneous specimens with varying dielectric constants, the dielectric behavior of such systems can be represented to arise from a combination of two lossy dielectrics in series. This internal barrier layer capacitance (ILBC) dominates the measured dielectric permittivity at higher temperatures and lower frequencies \cite{cite26,cite27}. At lower temperatures (and higher frequencies), the relative contribution from the layer capacitance reduces, allowing one to measure the contribution arising from the bulk \cite{cite28}. In our Bi$_2$CuO$_4$ specimen, we clearly observe two distinct features arising from these different contributions. The influence of the ILBC is observed in measurements with frequencies varying from 1 Hz to 400 Hz, above which this feature clearly moves to temperatures beyond 300K whereas at lower temperatures (and higher frequencies) the contribution arising from the grain can be easily discerned. We note that the value of the loss tangent is less than 1 for exciting frequencies in excess of 100 Hz, which makes Bi$_2$CuO$_4$ a more attractive candidate for applications in comparison with most other rare earth cuprates.  
 \begin{figure}
 	\centering
 	\hspace{-0.5cm}
 	\includegraphics[scale=0.32]{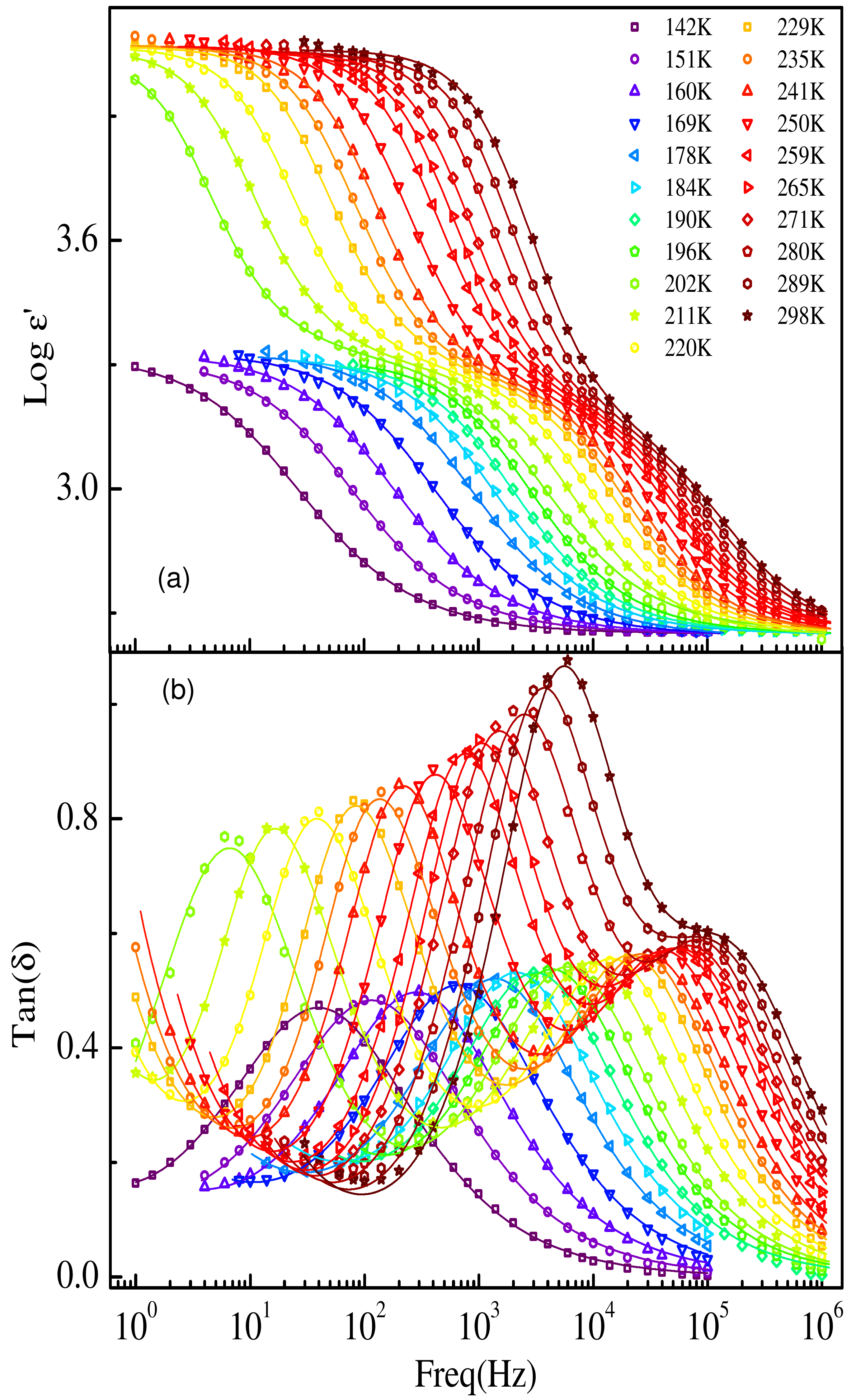}
 	\caption{The real part of the dielectric permittivity (a) and loss tangent (b) plotted as a function of the probed frequency at different temperatures. Solid lines are the fits according to Havriliak-Negami formulation.}
 	\label{Fig8}
 \end{figure}
\begin{figure}
 	\centering
 	\vspace{-0.2cm}
 	\hspace{-0.5cm}
 	\includegraphics[scale=0.38]{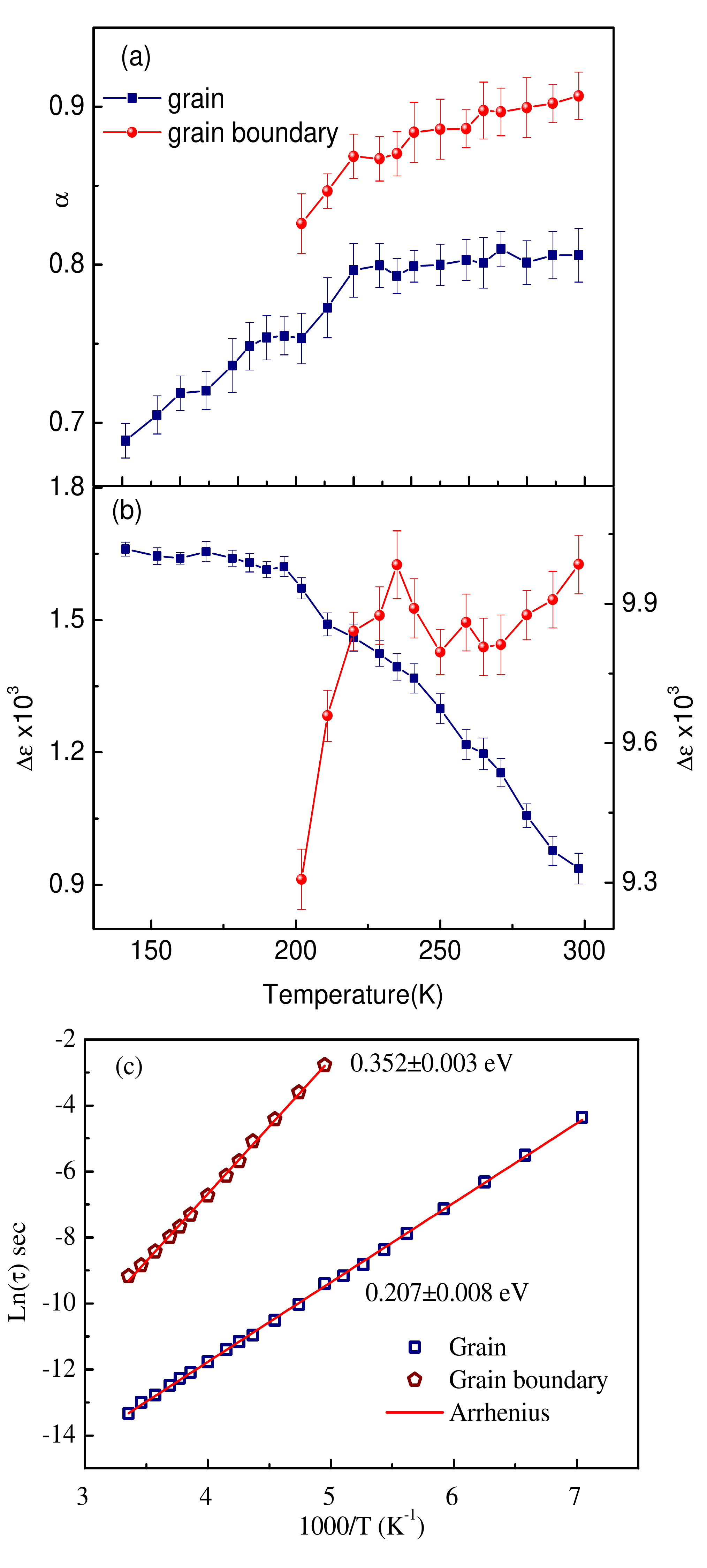}
 	\caption{The temperature dependence of (a) the broadening parameter $\alpha$ and (b) the dielectric strength $\Delta\epsilon$ as deduced from our fitting procedure for both the grain and the grain boundary contributions. (c) depicts the relaxation time ($\tau$) for each of these contributions fitted to an Arrhenius-like activated behavior.}
 	\label{Fig9}
 \end{figure}

These separate contributions to the effective dielectric constant can also be observed when the dielectric data is plotted in the form of Cole-Cole plots, with the frequency dependent imaginary part of the dielectric constant $\epsilon''$($\omega$) being plotted as a function of the real part $\epsilon'$($\omega$) at different temperatures. Fig. \ref{Fig7} depicts the Cole-Cole plots as measured in Bi$_2$CuO$_4$ in three different temperature regimes. At 150 K, a single semicircle is seen, indicating the presence of a single relaxation process  which in this case arises from the contribution of the grains. At 200 K on the other hand, two distinct relaxation processes are observed, with the high frequency semicircle corresponding to the intrinsic (grain) contribution, and the low frequency contribution arising from grain boundaries. At even higher temperatures (225 K), an additional upturn is observed at lower frequencies, which is a signature of the electrode polarization effect which presumably arises as a consequence of the accumulation of charges (the so called space-charge region) at the metallic electrodes. The presence of two relaxation processes is also clearly reflected in the frequency-domain data as is shown in Fig. \ref{Fig8}. In order to deduce quantities of interest, like the relaxation times, dielectric strengths and the peak parameters corresponding to the grain and grain boundary contributions, we have fitted the dielectric data with a generalized Hivriliak-Negami (H-N) function using the WinFit software from Novo-Control GmbH. Here, the frequency dependent complex dielectric permittivity is  given by 
\begin{equation} 
\begin{split}
\epsilon^{*} (\omega) = -i \ \Bigg( \frac{\sigma_0}{\epsilon_o\omega}\Bigg)^{N} +\sum_{k=1}^{2} \Bigg[\frac{\Delta\epsilon_k}{\bigg( 1+(i\omega\tau_k)^{\alpha_k}\bigg)^{\beta_k}}+\ \epsilon_{\infty k}\Bigg] \nonumber
\end{split}
\end{equation}
where $\omega=2\pi f$  is the probed frequency, $\epsilon_o$ is the permittivity of free space, $\tau$ is the associated relaxation time, $\sigma_o$ is the conductivity term resulting in the low frequency behavior of $\epsilon^{''}$, and N is the slope of $\epsilon^{''}(\omega)$ at lower frequencies ( $\approx$ 1 in most cases). The dielectric strength is denoted by $\Delta\epsilon = \epsilon{_S} - \epsilon{_\infty}$, where $\epsilon{_S}$ and $\epsilon{_\infty}$ refer to the low and high frequency dielectric permittivity. $\alpha$ and $\beta$ are broadening and asymmetry parameters linked to the relaxation peaks respectively, and can take values between 0 and 1. The H-N function modifies to a standard Debye function for $\alpha = \beta =$1. On the other hand, it modifies to a Cole-Cole function for $0<\alpha<1$ and $\beta =$1; and to a Cole-Davidson one for $\alpha =$1 and $0<\beta<1$. The fits to our frequency domain data are depicted in Fig. \ref{Fig8}, and both the real permittivity and the loss tangent is seen to fit well to the Cole-Cole functional form. Fig. 9a depicts the temperature dependence of the broadening parameter $\alpha$ as determined by our fitting procedure for both the grain and grain boundary contributions. The increase in the value of $\alpha$ with increasing temperature can be ascribed to the recovery of a Debye like behavior of non-interacting dipoles when the effective co-operativity is reduced due to thermal fluctuations. On the other hand, the temperature dependence of the dielectric strength ($\Delta\epsilon$) determined from our fitting is seen to be very different for the grain and grain boundary contributions, as is seen in Fig. 9b. In the high temperature regime, a para-electric like decrease of the dielectric strength of the grain is observed as a function of increasing temperature.  However, with increasing temperatures, the $\Delta\epsilon$ for the grain boundary contribution is seen to initially increase and then settle into a relatively $T$ independent value. The latter can be explained by invoking a scenario where increasing $T$ liberates more free charge carriers which are stuck at the grain boundaries, thus contributing to the extrinsic polarization. The relaxation times for both the grain and grain boundary processes have been calculated, and they are seen to confirm to an Arhenius type of activated behavior as is depicted in Fig. 9c. Data was fit using $\tau  = {\tau_0}  \exp \frac{E}{k_BT}$, where $\tau$, $\tau{_0}$, $E$ and $k_B$ refer to the relaxation time, pre-exponential factor, the activation energy, and the Boltzmann constant respectively. The linear fit shown in the inset provides activation energies of 0.207$\pm$0.008 eV and 0.352$\pm$.003 eV for the grain and grain boundaries respectively. Our observation that the activation energy of the grain-boundary is larger than that of the grain is consistent with that reported for other ceramic systems  \cite{cite23,cite26}. 

\section{Conclusion}
   In summary, we have investigated the cuprate Bi$_2$CuO$_4$ using magnetic, dielectric and pyroelectric measurements. We observe that a robust ferroelectric state is established in the vicinity of the magnetic transition, making this system a hitherto unreported exchange striction driven improper multiferroic. Magnetic and dielectric measurements reveal a regime of short range magnetic and polar order at $T > T{_N}$. At higher temperatures we find typical signatures associated with colossal dielectric constant materials, along with two distinct grain and grain boundary non-Debye type relaxations.  
\section{Acknowledgements}
The authors thank A. M Awasthi for extending experimental facilities for dielectric measurements, and for a critical reading of the manuscript. J.K. acknowledges DST India for a SERB-NPDF. S.N. acknowledges DST India for support through grant no. SB/S2/CMP-048/2013. The authors acknowledge funding support by the Department of Science and Technology (DST, Govt. of India) under the DST Nanomission Thematic Unit Program
\bibliography{Bibliography}

\end{document}